# Crystal growth, characterization and electronic band structure of TiSeS

Y. Shemerliuk[1], A. Kuibarov[1], O. Feia[1,3], M. Behnami[1], H. Reichlova[1&2], O. Suvorov[1], S. Selter[1], D.V. Efremov[1], S. Borisenko[1], B. Büchner[1&2], S. Aswartham[1]

[1] *Institut für Festkörperforschung, Leibniz IFW Dresden, Helmholtzstraße 20, 01069 Dresden, Germany*

[2] *Institut für Festkörper- und Materialphysik and Würzburg-Dresden Cluster of Excellence ct.qmat, Technische Universität Dresden, 01062 Dresden, Germany*

[3] *Kyiv Academic University, 03142, Kyiv, Ukraine*

## Abstract

Layered semimetallic van der Waals materials $TiSe_2$ has attracted a lot of attention because of interplay of a charge density wave (CDW) state and superconductivity. Its sister compound $TiS_2$, being isovalent to $TiSe_2$ and having the same crystal structure, shows a semiconducting behavior. The natural question rises - what happens at the transition point in $TiSe_{2-x}S_x$, which is expected for x close to 1. Here we report the growth and characterization of TiSeS single crystals and the study of the electronic structure using density functional theory (DFT) and angle-resolved photoemission (ARPES). We show that TiSeS single crystals have the same morphology as $TiSe_2$. Transport measurements reveal a metallic state, no evidence of CDW was found. DFT calculations suggest that the electronic band structure in TiSeS is similar to that of $TiSe_2$, but the electron and hole pockets in TiSeS are much smaller. The ARPES results are in good agreement with the calculations.

# 1. Introduction

In recent years, research on functional two-dimensional (2D) materials has stimulated activities aimed at synthesizing new materiaAls with novel functional properties. Due to their unique electronic [1-3], magnetic [4-6], and optical properties [7-9], the new 2D materials are expected to have great potential for applications. One of the most interesting class is the transition metal dichalcogenides $MX_2$ ($M$ = transition metal, $X$ = chalcogenide). They and their mixed systems dominate among the current 2D materials due to their favorable structural [10, 11], optical, electronic behavior [12-16].

$MX_2$ could be stabilized in a trigonal 1T, a hexagonal 2H, or a rhombohedral 3R structural phase. The electronic properties of $MX_2$ range from semiconductors [17] to semimetals [18, 19-20], from material with a trivial electronic structure till topological semimetals [21] and topological insulators [22]. Due to these unique properties, $MX_2$ continue to attract the scientific community.

TiSeS is a member of the trigonal 2D $MX_2$ family with the space group of *$P3\bar{m}1$(No. 164)*, where Ti atoms are sandwiched between two layers of S and Se atoms and each TiSeS layer is bound by the van der Waals (vdW) interaction. Therefore, the individual layers can be easily exfoliated. One of two pristine compounds of the mixed crystal system $TiSe_{2-x}S_x$ ($0 \leq x \leq 2$) considered with $x = 2$ is a semiconductor with an indirect band gap, while another with $x = 0$ is a semimetal with slightly overlapping bands [23]. In $TiSe_2$ compounds, a charge density wave (CDW) transition is observed at about 200 K at ambient pressure [24, 25]. $TiSe_2$ grown using high pressure shows insulating behavior at low temperatures [26]. However, Cu-intercalation or application of hydrostatic pressure suppress the CDW state, giving room for superconductivity [27, 28]. It was shown that the maximum superconducting $T_c$ corresponds to the CDW quantum critical point [29].

Motivated by this nontrivial behavior, in our work we try to use chemical pressure by substitution of Se by S to tune the material to the CDW quantum critical point. Since the ionic radius of S is smaller than Se, the substitution acts as an external pressure. As the first step, we optimized the synthesis and crystal growth conditions of TiSeS.. Further, we have investigated the crystal structure, magnetic and transport properties of as grown single crystals. Finally, we performed the band structure calculations using DFT and compare it with the ARPES measurements.

## 2. Experimental methods

The crystals were obtained from the Chemical Vapor Transport (CVT) crystal growth experiment. The chemical composition of our crystals was investigated using X-ray energy dispersive x-ray spectroscopy (EDX), with an accelerating voltage of 30 kV. Electron microscopic images were obtained by using a scanning electron microscope with two types of signals: the secondary electrons (SE) for topographic contrast and the backscattered electrons (BSE) for chemical contrast.

The crystal structure was investigated by powder X-ray diffraction (pXRD) using a STOE powder laboratory diffractometer in transmission geometry with Cu-K$_{\alpha 1}$ radiation (the wavelength ($\lambda$) is 1.540560 Å) from a curved Ge (111) single crystal monochromator and detected by a MYTHEN 1K 12.5° linear position sensitive detector manufactured by DECTRIS. An XRD pattern of a polycrystalline sample was obtained by grinding as-grown single crystals. Temperature and field dependent magnetization were measured on bulk as-grown crystal using a Quantum Design Superconducting Quantum Interference Device Vibrating Sample Magnetometer (SQUID-VSM). Transport measurements were carried on single crystals and the wires were glued by a conductive two component epoxy. Four contacts at the top surface *(ab-*plane) and two contacts on the bottom surface of the sample were prepared to allow for four probe method in both [in plane direction] and [c direction]. The measurements were performed in Oxford 15 T cryostat and the temperature was controlled by a heater at the sample holder. The electric current was applied either in the [*ab* plane] or [*c* direction], longitudinal and transversal voltage was measured. Magnetic field was applied perpendicular to the *ab*-plane for the charge carrier density measurements from Hall effect, the transversal data were antisymmetrized prior the mobility evaluation. Several samples were prepared showing similar results. ARPES measurements of TiSeS were performed on 1$^2$ station at BESSY synchrotron with Scienta Omicron R8000 energy analyzer with total energy and momentum resolution less than 10 meV and 0.01 Å$^{-1}$ respectively. High quality single crystals were cleaved and measured in a chamber with pressure better than 8*10$^{-11}$ mbar at a temperature less than 15 K.

**Crystal Growth via Chemical Vapor Transport:** TiSeS single crystals were grown by the chemical vapor transport technique. All preparation steps were performed under argon atmosphere in a glovebox, before sealing the ampule. The starting materials titanium (powder, Alfa Aesar, 99,99%) selenium (pieces, Alfa Aesar, 99.97%) and sulfur (pieces, Alfa Aesar, 99.999%) were weighed out with a molar ratio of Ti:Se:S = 1:1:1 and homogenized in an agate mortar. 1g of the starting material was loaded in a quartz ampoule (10mm inner diameter, 3mm wall thickness) together with 0.04 g of the transport agent iodine. The filled ampoule was cooled by liquid nitrogen, evacuated to a residual pressure of $10^{-8}$ bar and sealed at a length of approximately 12 cm by the oxy-hydrogen flame under static pressure.

The closed ampule was heated in a two-zone furnace with the following temperature profile optimized by us. Initially, the furnace was heated homogeneously to 790$^0$C with 100$^0$C/h. After that, an inverse transport gradient is applied to transport particles to the one side of the ampoule which is the charge region. This region was kept at 790$^0$C for 336 h. The other side of the ampoule which is the sink region was initially heated up to 840$^0$C at 100$^0$C/h, and dwelled at this temperature for 24 h. After that, the temperature in the growth zone was gradually reduced during one day to 730$^0$C to slowly form the transport temperature gradient for controlling nucleation and held at this temperature for 402 h. As a result, the temperature gradient was set for vapor transport between 790$^0$C (charge) and 730$^0$C (sink) for 17 days. Finally, the charge region was cooled to the sink temperature in 2 hours before both regions were furnace-cooled to room temperature.

Thin lustrous plate-like crystals of TiSeS perpendicular to the *c*-axis in the size of approximately 3 mm × 3 mm × 100 μm were obtained. As example, as-grown single crystals are shown in Fig. 1(a-b). All of these crystals show a layered morphology and they are easily exfoliated by scotch tape.

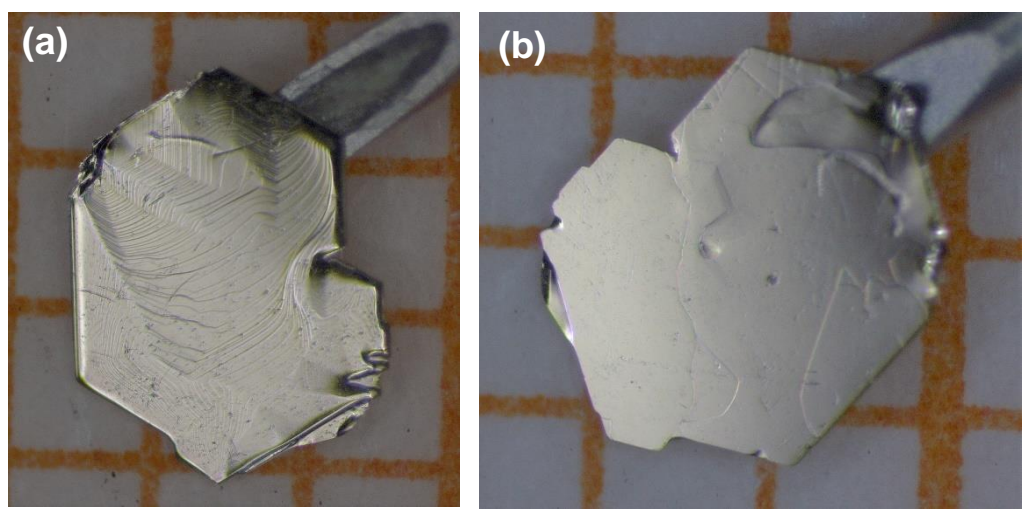

Figure 1 (a) and (b): As-grown crystals of TiSeS by the chemical vapor transport, cell scale is 1 mm

# Characterization: compositional and structural analysis

As-grown single crystals exhibit the typical features of layered systems, such as steps and terraces, as shown in Fig. 1(a-b). The topographical SE image of TiSeS crystal has a well-defined flat hexagonal facade with angles of 120°, which clearly indicates that they grew along the symmetry axes (Fig 2(a)). Back scattered electron (BSE) image of our crystal has homogeneous chemical contrast over the surface of the crystal, as shown in Fig. 2(b). This indicates a homogeneous elemental composition on the respective area of the crystal. At some small areas, the observed contrast changes can be clearly attributed to some scratches on top of the crystal and not to compositional changes, as clearly seen by comparing with the SE image.

The elemental composition of the TiSeS single crystals was determined by energy-dispersive x-ray spectroscopy (EDX) via measuring different areas and points on the surface of the crystal. The compositional analysis of as-grown single crystals is $Ti_{34.4(1)}Se_{30.2(9)}S_{35.3(2)}$. The compound shows the expected composition within the error of this measurement technique. The result of EDX measurements highly depends on the sample topography and the error regarding each element corresponds to an order of up to 5 at% [30].

The structural characterization and phase purity were confirmed by powder x-ray diffraction using a STOE powder diffractometer. The pXRD pattern obtained from TiSeS crystals, as shown in Fig. 3, was indexed in the space group $P\bar{3}m1$ *(No. 164)*, in agreement with literature [31]. No additional reflections were observed demonstrating the phase purity of our crystals. Starting from the crystal structure model proposed by Bozorth [32], a refined crystal structure model is obtained using the Rietveld method. Figure 3 together with the calculated pattern based on the Rietveld analysis (black line), the difference between measured and calculated pattern (blue line), and the calculated Bragg positions for a trigonal unit cell. The calculated peak positions and intensities are in satisfactory agreement with the experimental data. TiSeS has a $CdI_2$-type structure with the titanium atom at (0, 0, 0) position and the chalcogen atoms at (1/3, 2/3, z) position with the random distribution of S and Se on a position Fig. 4. The obtained lattice parameter and reliability factors are summarized in Tab. 1 and are illustrated in Fig. 3.

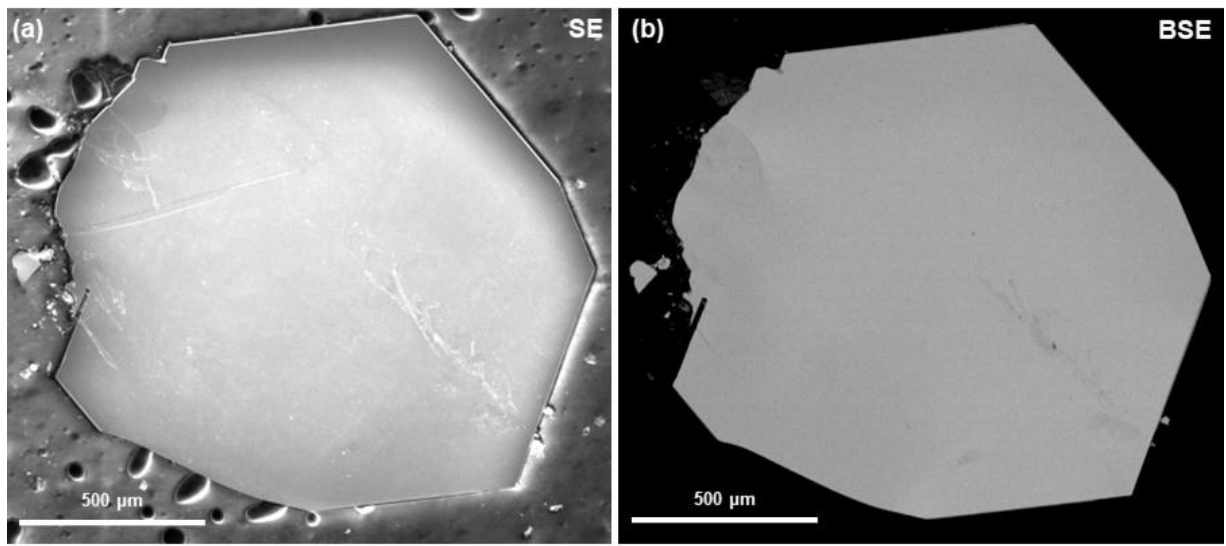

Figure 2. SEM image of an as-grown TiSeS crystal with topographical contrast (SE mode) in (a) and chemical contrast (BSE mode) in (b).

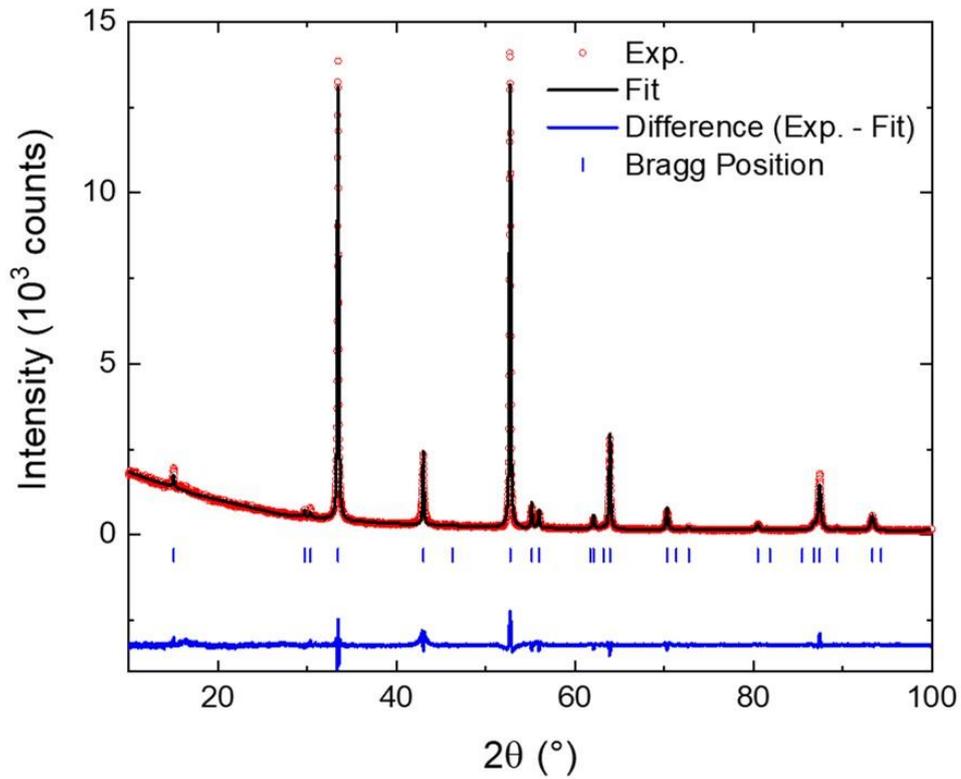

Figure 3. Powder x-ray diffraction pattern together with the Rietveld analysis of TiSeS.

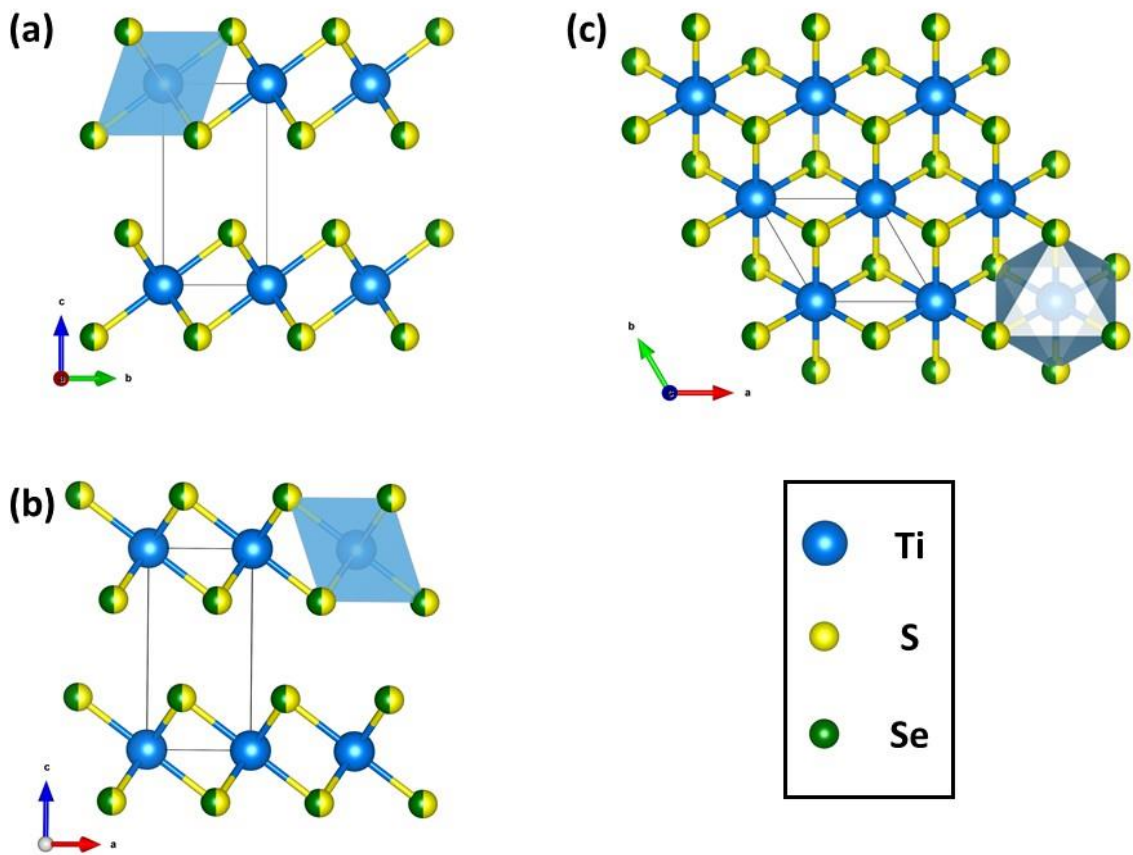

Figure 4. As an example crystal structure of TiSeS (a) shown in the *bc* plane, (b) shown in the *ac* plane and (c) shown in the *ab* plane. The graphical representation was prepared using VESTA3 [33].

.

|             | pXRD          |
|-------------|---------------|
| *Composition* | TiSeS       |
| *Space group* | P$\bar{3}$m1(No. 164) |
| *Wavelength (Å)* | 1.540560 |
| *2θ range (°)* | 10 – 100   |
| *Step Size (°)* | 0.015     |
| *Temperature (K)* | 293     |
| *a* (Å)     | 3.4681(8)     |
| *c* (Å)     | 5.879(3)      |
| *V* (Å$^3$) | 61.243(3)     |
| *Goodness-Of-Fit* | 2.67    |
| *Bragg R-factor* | 6.372    |
| *RF-factor* | 6.967         |

Table 1. Structural parameters and residual factors of Rietveld refinement

## 3. Magnetization and transport
### Magnetic properties:

The magnetic properties of the as-grown single crystals as well as polycrystalline samples of TiSeS were measured by SQUID-VSM. The measurements were performed both *H//ab & H//c*. The magnetization as a function of temperature is shown in Fig. 5(a), the magnetization curve shows paramagnetic like behaviour with slight negative values for *H//ab* which might be coming from excess amount of diamagnetic epoxy used to mount the crystal. To improve the measurement parameters and subsequent calculations, an additional measurement of the polycrystalline sample was made without the use of epoxy.

As shown in Fig. 5(b) the molar susceptibility as a function of temperature shows paramagnetic behavior without any long range order. Inverse susceptibility shown in Fig. 5(c). reveals that susceptibility becomes non-linear below ~250K, i.e. paramagnetic-like behavior is observed only at relatively high temperatures. Linear fit of the Curie-Weiss law results in theta $\theta = -282K$ with an effective moment $\mu = 0.83\ \mu_B$, per unit cell. It indicates of frustrated AFM-like interaction. At temperatures below 200 K we observe deviation from the Curie-Weiss law. The high effective moment has also observed in the Ti-vacant samples in TiSe$_2$ parent compound [34] they concluded that localized spins Ti at the interstitial positions or vacancies of Ti. Here also we observe similar high effective moment in TiSeS, most likely the localised moments of Ti-are present. This is highly interesting as the localized spins can give interesting behavior such as spin glass, cluster glass [34-36]. In another recent work on [37], non-magnetic system, TiO$_2$ also exhibit large moments that can be caused by defects in the crystal structure.

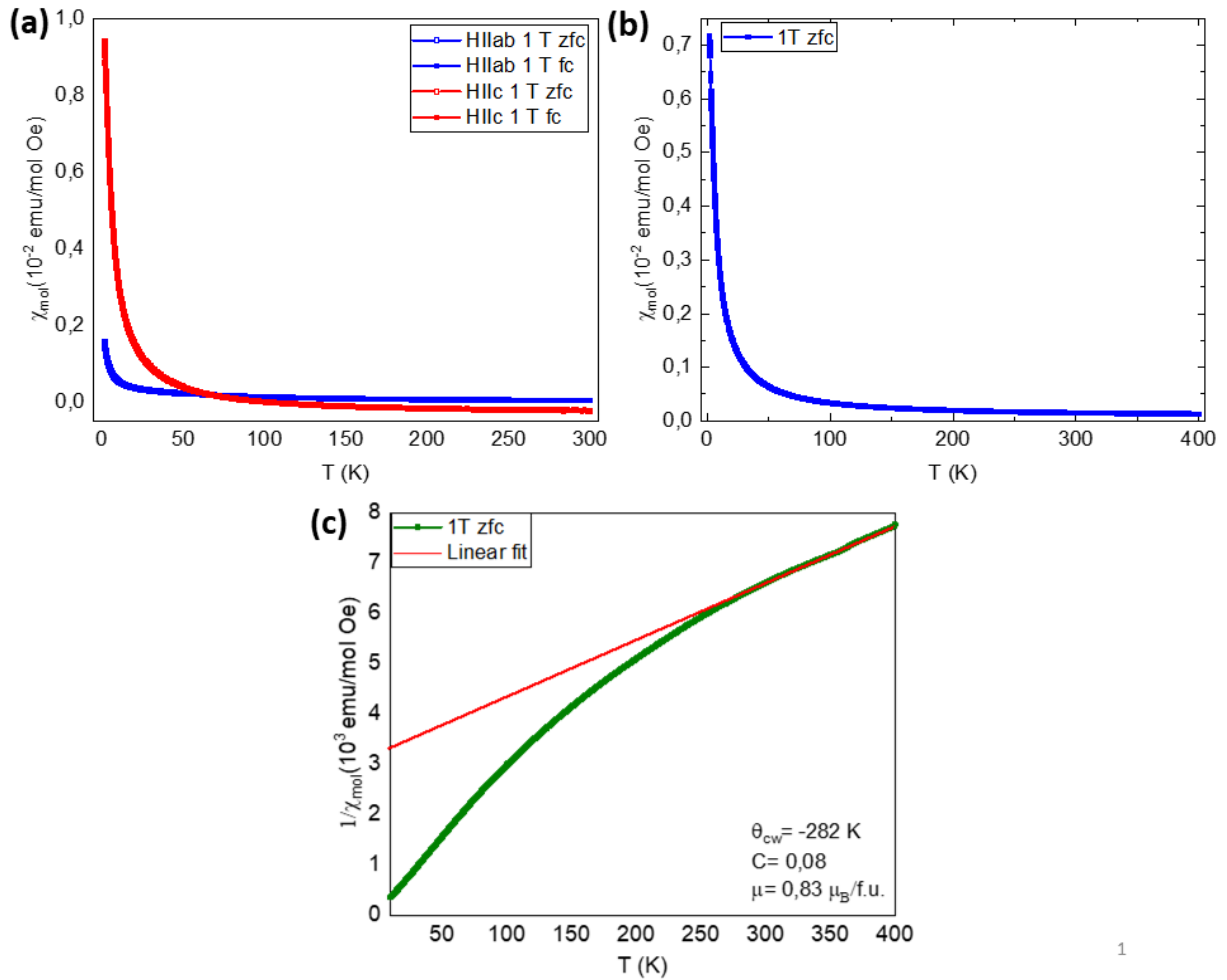

Figure 5 (a) Molar susceptibility as a function of temperature for a field of 1 T applied *H//ab* and *H//c* for TiSeS crystal (b) molar susceptibility as a function of temperature for a field of 1 T for TiSeS polycrystal (c) inverse molar susceptibility temperature dependence with Curie-Weiss linear fit

## Transport properties:

Electrical resistivity measurement shows strongly anisotropic metallic behavior; $\rho_{xx} \sim T^n$ with n = 1.7 at low temperature. No indications of the critical fluctuations were observed down to 5 K suggesting that we are not in the QCP part of the phase diagram. The sheet resistivity is monotonously decreasing with temperature unlike [38] showing metal like behavior in the studied temperature range (Fig. 6(a)). The resistivity was measured with electrical current applied along various crystal directions: in the *ab*-plane and in the *c* direction. The relative change of the resistivity with temperature is larger when electric current is applied in the *ab*-plane pointing to the anisotropic transport properties arising from the van der Waals character of the material.

The magneto-resistance at various temperatures was measured at longitudinal contacts while sweeping magnetic field in the *c* direction as shown in Fig. 6(b). At low temperatures the magneto-resistance is increased reaching 0.3% at 10 K. Interestingly, it deviates from the ordinary quadratic magnetoresistance and it has a negative character. It is probably caused by weak-localization effects due to the low dimensionality, as observed previously in van der Waals systems [39].

The carrier density concentration was estimated from the Hall effect (Fig. 6(c)) at various temperatures. The mobility at 10 K is 102 cm$^2$/V.s and decreases at 100 K to 76 cm$^2$/V.s. At low temperature a weak non-linearity in the Hall data with respect to the magnetic field was observed. The magnetic field can influence the complex band structure of the material leading to such non-linearity at high magnetic fields.

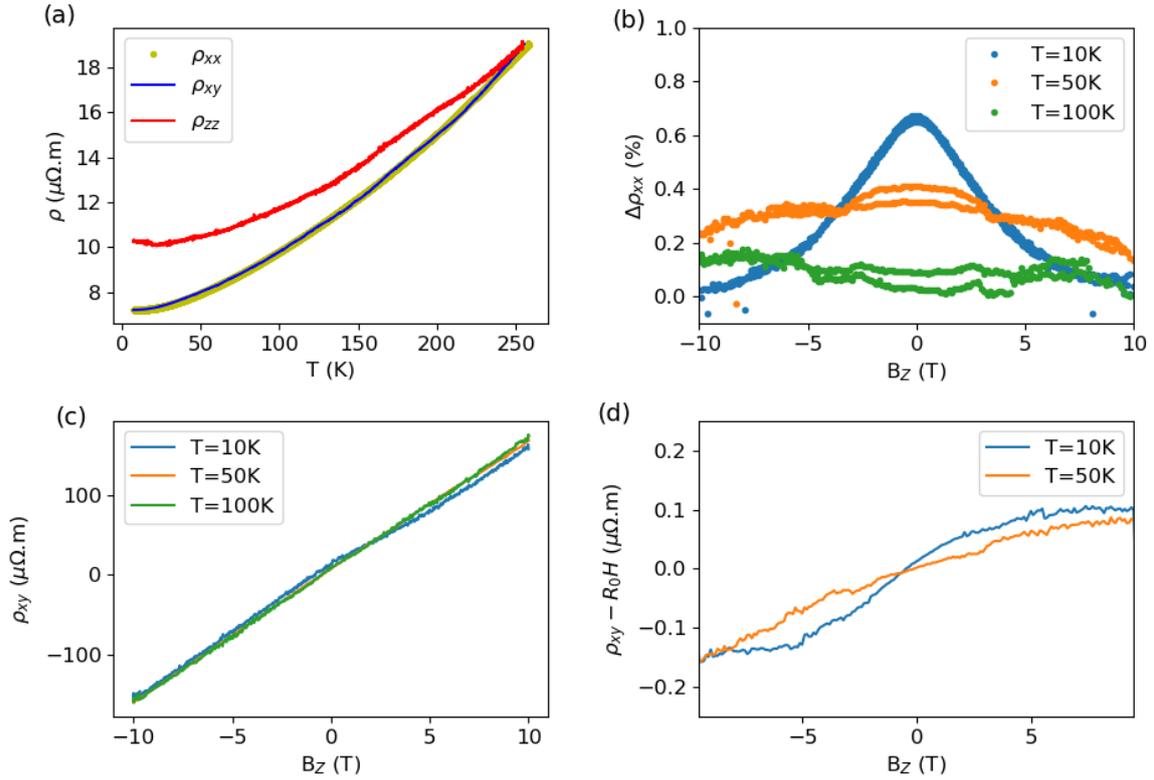

Figure 6. Magneto-transport characterization of TiSeS single crystals (a) Resistivity as a function of temperature for different crystal direction of the TiSeS single crystal. $\rho_{xx}$ and $\rho_{yy}$ were measured with electric current in the ab-plane and the $\rho_{zz}$ was measured with electric current applied along the *c*-direction. (b) Magneto resistance measured at longitudinal contacts at various temperatures when magnetic field is applied along the *c* direction. (c) Hall measurements at various temperatures. (d) Linear slope was subtracted from the measured Hall signal to highlight the non-linearity in magnetic field.

## 4. DFT-Calculation:

Using *ab initio* calculations, we investigate the electronic band structure of $TiSe_2$, $TiS_2$ and TiSeS as well as the possible crystal structures. To investigate possible crystal structures of the new TiSeS crystal, we have performed an evolution search in conjunction with ab initio calculation in the frame of the density functional theory (DFT). For structure prediction of systems with mixed chalcogen composition we used USPEX code [40, 41]. One can explore low-energy configurations for chosen composition with technique implemented in this code. To predict thermodynamic stability of the ternary crystals we used non-variable search mode with minimum 12 atoms and maximum 18 atoms for 1st generations. It was produced by random structure generator and consisted of 100 structures. Each of the following generations had 8 compositions, which was obtained by applying 40% heredity, 20% softmutation, 10% lattice mutation and 30% random structure generator. To check possible low-energy compositions we had performed calculations for ternary stoichiometries within Ti and S and Se atoms at ratio 1:1:1. Each structure was carefully relaxed in four stages, starting from low precision.

For structure relaxations within USPEX we used DFT [42] implemented in the VASP code [43-44]. Perdew-Burke-Ernzerhof (PBE) generalized gradient approximation (GGA) [45] and projector-augmented wave (PAW) [46, 47] within DFT was used for these calculations. We have set the wave kinetic energy cutoff to 350 eV, and uniform Γ-centered k-meshes with reciprocal-space resolution of $2\pi \times 0.07$ Å$^{-1}$ were used for Brillouin zone sampling. In post-processing we carefully relaxed all low-energy structures in USPEX output with 12x12x6 k-points mesh and energy cutoff 400 eV.

As the result of evolutionary search, we have got the set of low energy structures within Ti-Se-S system. The most stable structure has 156 point symmetry, and it consists of three different atom layers – Ti in the middle, while Se and S atoms are located on up and down layers. It is known that PBE approximation does not correctly treat van-der-Waals forces. To check if this material corresponds to experimental finding of this paper, we have applied dispersion correction to our computational finding with Becke-Jonson dumping PBE-D3(BJ) [48, 49] and the Grimme zero dumping [49] PBE-D3(ZD). For comparison we have applied the same procedure to TiS$_2$ and TiSe$_2$ structures.

| Lattice constant | Experimental | PBE | PBE-D3(BJ) | PBE-D3(ZD) |
|---|---|---|---|---|
| TiS$_2$ | | | | |
| a | 3.407 | 3.409 (~0%) | 3.368 (-1.14%) | 3.400 (-0.21%) |
| c | 5.695 | 6.487 (+14.08%) | 5.542 (-2.68%) | 5.687 (-0.14%) |
| TiSeS | | | | |
| a | 3.4681(8) | 3.458 (-0.29%) | 3.408 (-1.74%) | 3.444 (-0.70%) |
| c | 5.879(3) | 6.743 (+14.69%) | 5.789 (-1.53%) | 5.984 (+1.78%) |
| TiSe$_2$ | | | | |
| a | 3.540 | 3.534 (-0.17%) | 3.479 (-1.72%) | 3.520 (-0.56%) |
| c | 6.010 | 6.736 (+12.08%) | 5.895 (-1.91%) | 6.065 (+0.92%) |

Table 2. Structural parameters (in Angstrom) for Ti-Se-S system obtained with PBE pseudopotentials with and without disperse correction

Experimental data for TiS$_2$ and TiSe$_2$ we collected from Springer Materials database. From Table 2 one can see the common GGA-PBE method overestimates *c* lattice vector by 12-14% for all considered structures. While inclusion of dispersion correction gives more reasonable estimation by 1-2% difference with experimental results. Zero dumping method PBE-D3(ZD) gives the best fitness to experimental results with almost precise lattice constants length.

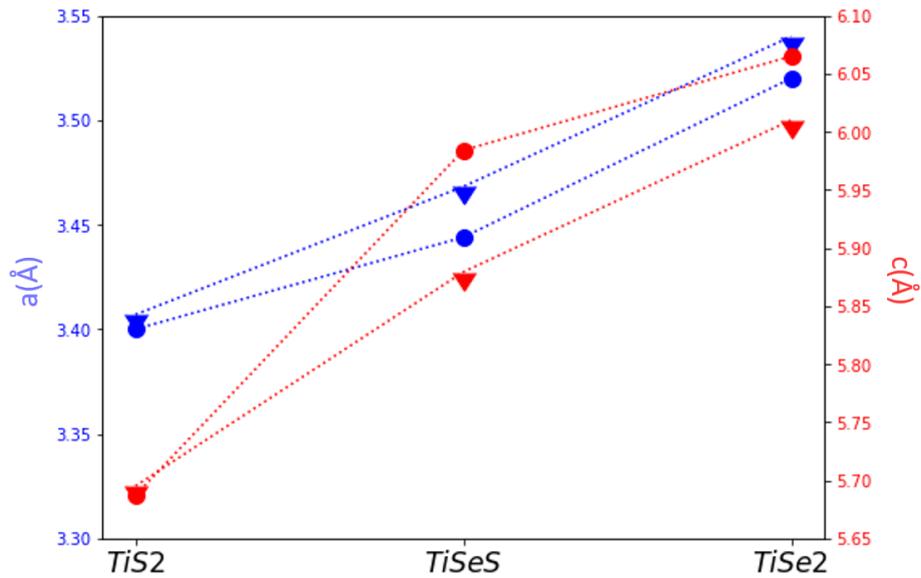

Figure 7. Lattice constants for TiX$_2$ (X = S, Se) system, obtained with PBE in conjunction with zero dumping dispersion correction. Triangles stands for experimental data, circles – for calculated parameters.

Unit cells of these structures expand in line from TiS$_2$ to TiSe$_2$ via TiSeS. As shown on Fig 7 lattice vectors *a* and *c* for TiSeS lays on a line between parameters for TiS$_2$ and TiSe$_2$. This can be explained by changing the chalcogen's size followed with changing the structural constants.

Low energy structures, obtained by evolutionary search within PBE approximation, are presented on Figure 8 and their structural parameters are collected in Table 3. Here we present part of results for structures within 20 meV/f.u (f.u = formula unit) from material with 156 symmetry. All of them share structural similarity – they consist of trilayers with Ti-layer as central element, but with different arrangement of Si and S atoms. The most stable in our calculations is the structure with the crystal structure space group #156-TiSeS (Fig. 8(a)), which consists of three layers S-Ti-Se. Structure #8-TiSeS (Fig. 8(b)) with huge unit cell with 4 trilayers shows some mixture of these atoms. It consists of S-Ti-Se trilayers, separated by trilayers with Se-S composition. Similar structure shares #6-TiSeS (Fig. 8(d)), while #1-TiSeS (Fig. 8(d)) has one trilayer in the unit cell, but expanded *a* lattice constant. Along *a* its chalcogen layers consist of 1 Se (1 S) atom and 3 S (3 Se) atoms. While it is not possible in experimental conditions to achieve ideal composition, we believe there should be a mixture of different trilayers in grown crystals. As all discussed structures are close in energy, their mixture should result into 164 symmetry, which is observed in experiment.

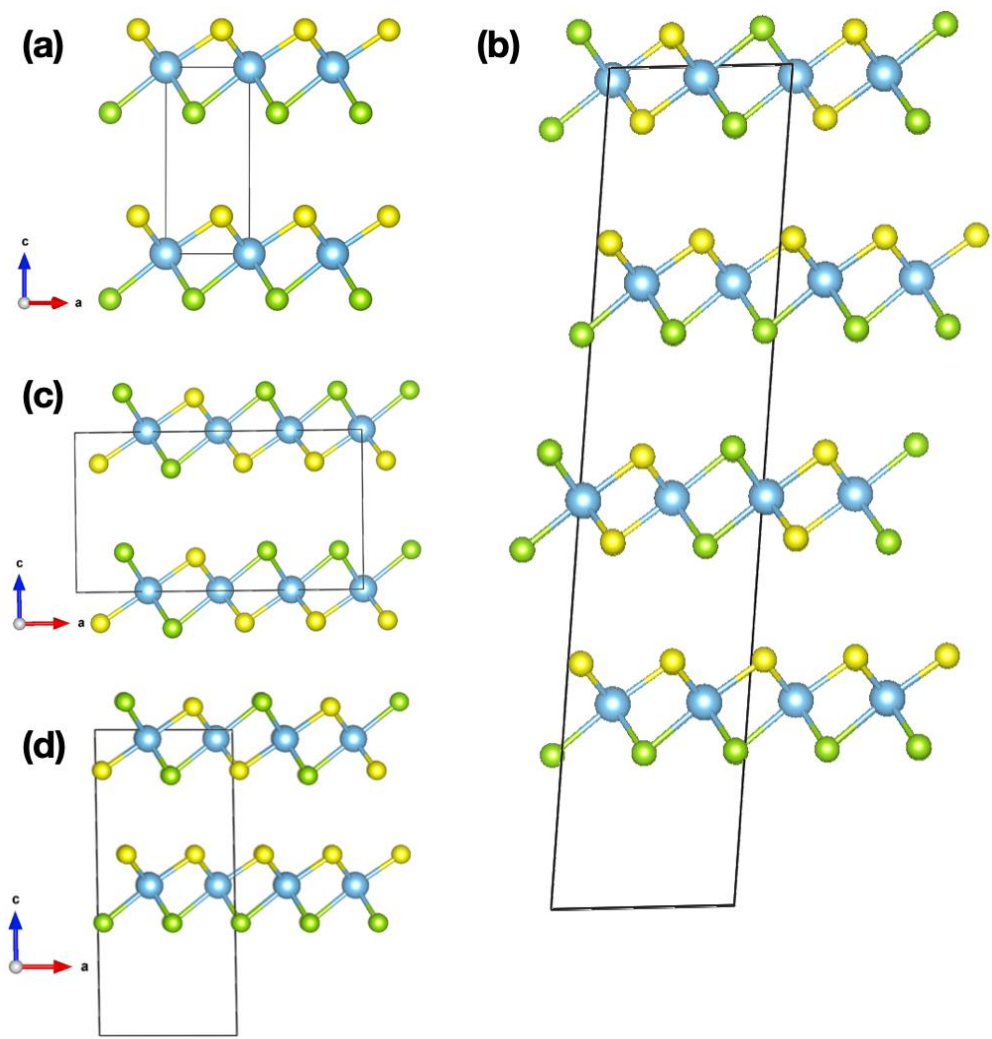

Figure 8. Low energy Ti-Se-S structures, obtained with evolutionary search: (a) 156 symmetry; (b) 8 symmetry; (c) 1 symmetry; (d) 6 symmetry. Blue balls indicate Ti atoms, yellow and green – S and Se atoms correspondingly

| Symmetry | Structure constants (A) | Angles | Energy difference (meV/f.u) |
|---|---|---|---|
| 156 | a= 3.458<br>c=6.740 | $\alpha = \beta = 90$<br>$\gamma = 120$ | 0 |
| 8 | a=3.470<br>b=5.979<br>c=27.793 | $\alpha = \gamma = 90$<br>$\beta = 95.204$ | 14 |
| 1 | a=11.991<br>b=3.465<br>c=6.741 | $\alpha = 90.644$<br>$\beta = 89.926$<br>$\gamma = 89.974$ | 15 |
| 6 | a=5.973<br>b=3.465<br>c=13.392 | $\alpha = \gamma = 90$<br>$\beta = 90.851$ | 18 |

Table. 3. Low energy Ti-Se-S structures, obtained with evolutionary search with PBE approximation – symmetry, lattice parameters and energy difference per formula unit

# 5. ARPES:

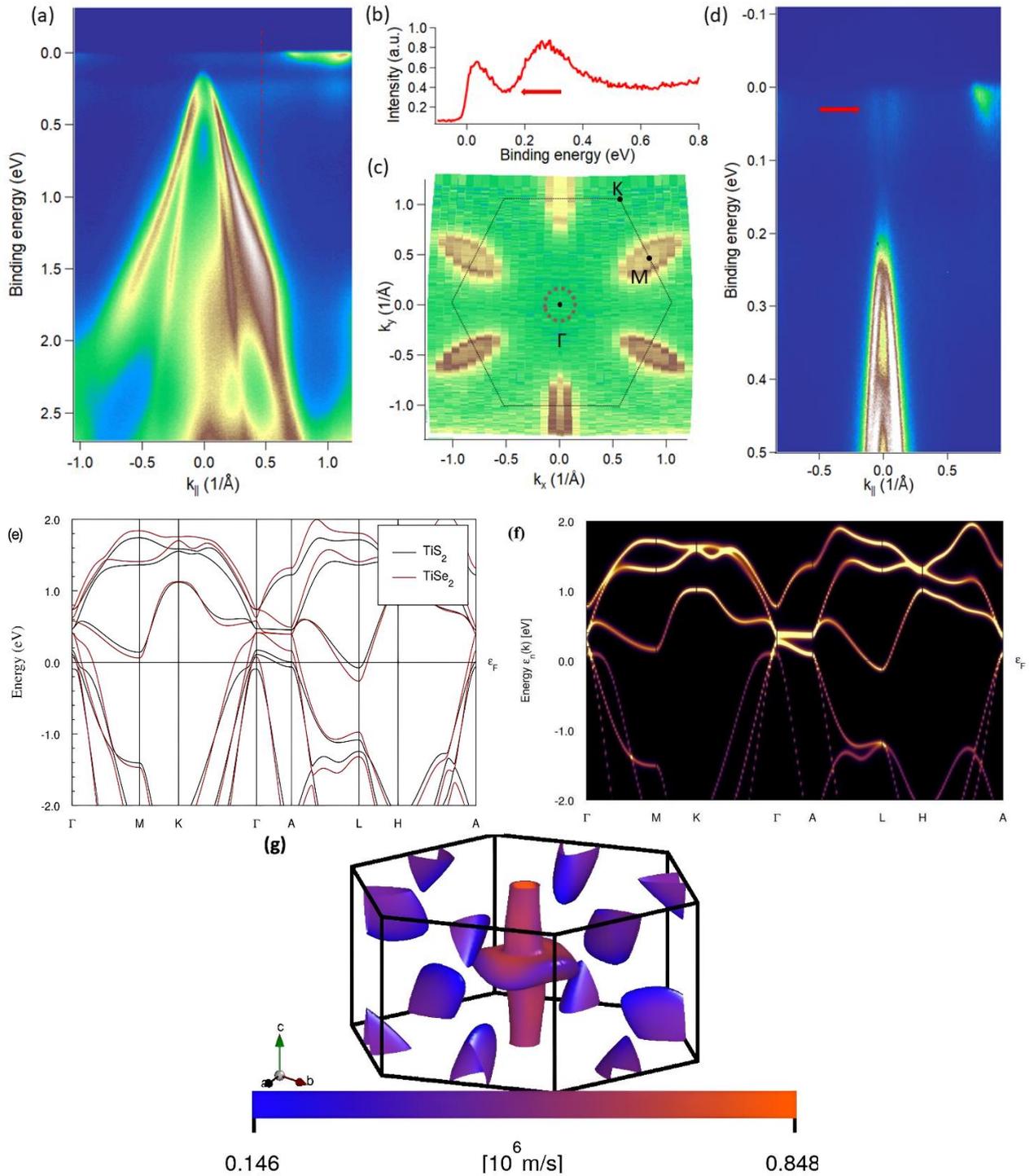

**Figure 9** ARPES spectra of TiSeS measured at 15 K along Γ-M direction using (a) 70 eV and (d) 35 eV photon energy. (b) Energy distribution curve (EDC) taken at $k_{\parallel}$ = 0.5 Å$^{-1}$. (c) Symmetrized Fermi surface map of TiSeS recorded with 80 eV photon energy. Brown dashed circle in center highlights electron pocket around Γ not clearly visible in the map due to strong intensity of electron pockets around M-points. The electron-like pocket around Γ is clearly seen in panel (d) where we increase the dynamic range of the color scale enhancing the weak features. Dotted black hexagon represents the Brillouin zone. (e) Calculated band structure of experimental TiSe$_2$. (f) Band structure spectral function in CPA approximation for TiSeS (g) Fermi surface of TiSe$_2$ with the structure parameters TiSeS, the color corresponds to the Fermi velocity

Measurements of TiSeS were performed on 1^2 station at BESSY synchrotron with Scienta Omicron R8000 energy analyzer with total energy and momentum resolution less than 10 meV and 0.01 Å$^{-1}$ respectively. High quality single crystals were cleaved and measured in a chamber with pressure better than $8*10^{-11}$ mbar at a temperature less than 15 K. ARPES spectra in Fig.9 show that TiSeS hosts almost the same electronic structure as its parent compound - TiSe$_2$. Three hole-like dispersions around $\Gamma$-point and one electron-like pocket around M-point. Measurements confirm previous statements as regards the absence of commensurate charge density waves (CDW) [50] in TiSeS. There are no visible umklapp hole-like bands near M-points. Figure 9(b) represents EDC taken at $k_{\parallel} = 0.5\pm0.02$ Å$^{-1}$ (red dashed line in Fig. 9(a)) and shows suppression of spectral weight around 0.14 eV binding energy (red arrow) visible in ARPES spectra which might be a sign of electron-phonon or electron-hole interaction that leads to formation of CDW in TiSe$_2$. In Figure 9(d) red arrow points to additional, comparing to TiSe$_2$, electron pocket in $\Gamma$-point which might be associated with the suppression of CDW state in this compound. Its spectral weight is too weak comparing to electron pockets in M therefore this band is almost not visible in Fermi map Fig. 9(c). Since the substitution of Se by S is isovalent, we performed the calculations for TiSe$_2$ and TiS$_2$ using the experimental structural parameters of TiSeS (Tab I). The obtained band structures are shown in Fig. 9(e). The electronic structures of both compounds are similar to the electronic structure TiSeS in the coherent potential approximation (CPA) which is shown in Fig. 9(f). The calculation was done within the package FPLO [51, 52]. In order to get clue of the Fermi surface we calculate TiSe$_2$ with the structure parameters of TISeS. The results are shown in Fig. 9(g). Comparison to ARPES (Fig. 9(c) of the Fermi surface show a good agreement.

## 6. Conclusions

We report the growth and electronic structure characterization of TiSeS single crystals. TiSeS crystals exhibit a layered morphology with weak van der Waals interactions between layers parallel to the crystallographic *ab*-plane of the trigonal structure in the space group $P\bar{3}m1$(No. 164). Transport measurements show negative magneto resistance stemming from weak-localization effects. No indications of CDW were found in the transport. Our experiments were supplemented by the electronic band structure calculations in the frame of the density functional theory (DFT). Electronic band structure in TiSeS is similar to that in TiSe$_2$. The main difference is that the electron and hole pockets are much smaller in TiSeS. The obtained band structure is confirmed in further angle-resolved photo emission (ARPES) experiments.

We have studied band structure and low energy metastable states in the frame of the DFT theory. The lowest energy state is the one where the layers are three separate layers S-Ti-Se. However, there are many other low energy states. The "averaging" leads to the more symmetrical state *No. 164* observed in the experiment.

## Acknowledgements

S.A. acknowledges the support of Deutsche Forschungsgemeinschaft (DFG) through Grant No. AS 523/4–1. S.A., D.E. acknowledge financial support DFG through the project 405940956. D.E. acknowledges VW foundation for support though the trilateral project "Synthesis, theoretical examination and experimental investigation of emergent iron-based superconductors". B.B. acknowledges financial support from the DFG through the Würzburg-Dresden Cluster of Excellence on Complexity and Topology in Quantum Matter – ct.qmat (EXC 2147, project-id 390858490). We thanks Ulrike Nitzsche for technical assistance. O.F. acknowledges UKRATOP project supported by BMBF and support from the National Research Foundation of Ukraine (Project 2020.02/0408)

y.shemerliuk@ifw-dresden.de

s.aswartham@ifw-dresden.de